\documentclass[twocolumn,showpacs,preprintnumbers,amsmath,amssymb,superscriptaddress]{revtex4-1}
\usepackage{graphicx}
\usepackage{dcolumn}
\usepackage{bm}
\usepackage{CJK, CJKnumb}
\usepackage{color}
\begin{document}
\begin{CJK*}{GBK}{song}

\title{Equal-spin and oblique-spin crossed Andreev reflections in ferromagnet/Ising superconductor/ferromagnet junction}
\author{Wei-Tao Lu} \email{Email address of Wei-Tao Lu: physlu@163.com}
\affiliation{School of Physics and Electronic Engineering, Linyi University, Linyi 276005, China}
\author{Qing-Feng Sun} \email{Email address of Qing-Feng Sun: sunqf@pku.edu.cn}
\affiliation{International Center for Quantum Materials, School of Physics, Peking University, Beijing, 100871, China}
\affiliation{Collaborative Innovation Center of Quantum Matter, Beijing 100871, China}
\affiliation{Beijing Academy of Quantum Information Sciences, West Bld. $\#$3, No. 10 Xibeiwang East Road, Haidian District, Beijing 100193, China}
\author{Qiang Cheng}
\affiliation{School of Science, Qingdao University of Technology, Qingdao 266520, China}

\begin{abstract}
We study the subgap transport through a ferromagnet/Ising superconductor/ferromagnet (F/ISC/F) junction by solving the Bogoliubov-de Gennes equations. It is found that the crossed Andreev reflection (CAR) and local Andreev reflection (LAR) strongly depend on the spin polarized F, the magnetization direction, and the Ising superconducting phase. For the same magnetization directions of the two F leads, the equal-spin CAR could take place due to spin-flip mechanism induced by the Ising spin-orbit coupling and equal-spin-triplet pairing. Both equal-spin CAR and equal-spin LAR exhibit a remarkable magnetoanisotropy with period $\pi$ and show oscillatory behavior with chemical potential. The equal-spin CAR is more prominent for half-metal F and double-band ISC while the normal CAR is completely suppressed. When the magnetization directions of the two F leads are different, the oblique-spin CAR occurs and its magnetoanisotropic period generally becomes $2\pi$ instead of $\pi$. In the oblique-spin CAR process, the spins of the electron and hole are neither parallel nor antiparallel. Furthermore, the property of oblique-spin CAR is very sensitive to the spin and valley degrees of freedom. The spin and valley polarized CAR can be achieved and controlled by the chemical potentials and the magnetization directions.
\end{abstract}
\maketitle

\section{Introduction}

Transition-metal dichalcogenides (TMDs) have attracted significant attention, which provide an ideal platform to realize the spintronics and valleytronics \cite{Wang, Schaibley, Manzeli}. TMDs are regarded as new two-dimensional semiconductors with a large direct band gap \cite{Mak}. The broken inversion symmetry and strong spin-orbit coupling (SOC) lead to a spin-valley locking relationship in TMDs, where the spin splitting of the valence bands is opposite at the two valleys due to the time-reversal symmetry \cite{Xiao}. Thus, the SOC of TMDs is also referred as the Ising SOC field \cite{Law1}. Furthermore, recent theoretical \cite{Law1, YGe, Roldan, Rosner, Yuan, CWang, Shaffer} and experimental \cite{JTYe, Taniguchi, Costanzo, JMLu, YSaito, XXi, Barrera, JMLu2} studies reveal that the TMDs, such as MoS$_2$, MoSe$_2$, MoTe$_2$, WS$_2$, and NbSe$_2$ could become superconducting. The gated MoS$_2$ thin films exhibit electric field-induced superconductivity with a critical temperature $10K$ \cite{JTYe, Taniguchi}. In particular, due to the Ising SOC where the spin is pinned to the out-of-plane direction, the in-plane critical field of the system is far beyond the Pauli limit \cite{JMLu, YSaito}. The superconducting monolayer NbSe$_2$ possesses an in-plane critical field of more than six times the Pauli limit, suggesting an unconventional Ising pairing protected by spin-momentum locking \cite{XXi, Barrera}. These works provide experimental evidence of an Ising superconductor (ISC) with spin and valley degrees of freedom in the two-dimensional where the spins of the pairing electrons are strongly pinned by Ising SOC \cite{JMLu, YSaito, XXi, Barrera, JMLu2}.

Andreev reflection (AR) is a process of electron-hole conversion which occurs at the interface of a normal metal (N) and superconductor (SC) \cite{Andreev, Beenakker, Sun1, Sun2}. The incident electron in the N is reflected at the interface as a hole and a Cooper pair is injected into the SC. Generally, the AR would be suppressed and the subgap conductance would be zero when the ferromagnet (F) is completely spin polarized in the F/SC junction. However, the interfacial Rashba SOC and spin-flip mechanism could lead to a nonzero AR even when the F is completely spin polarized \cite{Lofwander, Niu, Hogl, Beiranvand}. This anomalous AR is an equal-spin AR by equal-spin-triplet pairing in which an electron is reflected as a hole with the same spin. Alternatively, the ISC opens a new route for studying the equal-spin AR \cite{Law1, Sun3, Sun4}. Law et al. found that due to the equal-spin-triplet Cooper pairs generated by Ising SOC, the equal-spin AR can occur in the half-metal/ISC junction and Majorana fermions can be created in a half-metal wire placed on top of ISC \cite{Law1}. The equal-spin AR through the F/ISC junction exhibits a strong magnetoanisotropy and the magnetoanisotropic period is $\pi$ instead of $2\pi$ as in the conventional magnetoanisotropic system \cite{Sun3}. In ISC Josephson junctions, the switch effect of equal-spin Josephson current and $0-\pi$ transitions can be achieved due to the equal-spin AR and the Ising pairing order parameter \cite{Sun4}. Even so, the study on the equal-spin crossed Andreev reflection (CAR) in the F/ISC/F junction has yet to be explored.

CAR is a nonlocal AR process where the conversion from an electron into a hole occurs at two different interfaces of SC. As an application, the CAR is a good approach to produce the spatially separated entangled states of electrons by splitting Cooper pairs \cite{PRecher, JNilsson, Herrmann, Tan}. The experiments demonstrate that the Cooper pair splitters taking advantage of CAR could be realized in carbon nanotubes and graphene \cite{Herrmann, Tan}. Recently, many works on CAR have been reported in various systems, including graphene \cite{Cayssol, JLinder, Beiranvand2}, silicene \cite{JLinder2, HLi, WTLu}, TMDs \cite{LMajidi, CBai}, topological insulators \cite{Reinthaler, JWang, Crepin, Islam}, Weyl semimetals \cite{YLiu, XSLi, HLi2}, and topological superconductor \cite{YTZhang, YFZhou, QLi} and so on. The Rashba SOC could produce an anomalous CAR and negative charge conductance in the F/Rashba SOC/SC/F junction of graphene, which is linked to the equal spin triplet pairing \cite{Beiranvand2}. The topologically protected edge state in topological insulators provides a feasible opportunity to realize the robust CAR. Due to the helical edge states in toplogical insulator, there is a unique relation between the CAR process and the odd-frequency triplet superconductivity in F/SC/F junction \cite{Crepin}. Considering the valley and spin degrees of freedom, the valley/spin-selective CAR and valley/spin-entangled states have been carried out where the Cooper pair is composed of electrons from the opposite spins in opposite valleys \cite{JLinder2, HLi, WTLu, LMajidi, CBai}. Furthermore, the electrical control of CAR and spin-valley switch in antiferromagnet/SC/antiferromagnet junctions are reported very recently \cite{WTLu, Jakobsen}.

Motived by the works on ISC and CAR, in this paper, we study the CAR and local AR (LAR) in the F/ISC/F junction, as shown in Fig. 1(a). The results show that due to the spin flip originating from the Ising SOC, the equal-spin CAR could occur, the physical mechanism of which is different from that of normal CAR. The equal-spin CAR exhibits a strong magnetoanisotropy with period $\pi$ and could achieve its maximum for an in-plane magnetization direction. We demonstrate a significant increase of the equal-spin CAR for half-metal F. The spin and valley dependent oblique-spin CAR can be controlled by the magnetization direction and the Ising superconducting phase. The effect of structural parameters on the CAR and LAR is discussed in detail.

The rest of this paper is organized as follows. The Hamiltonian of the F/ISC/F junctions and the corresponding band structures are given in Sec. II. The results on equal-spin CAR, oblique-spin CAR, and equal-spin LAR are discussed in Sec. III. A brief summary is presented in Sec. IV.

\section{Theoretical Formulation}

The TMDs such as MoS$_2$ can be regarded as strongly bonded two-dimensional S-Mo-S layers. In spite of the complexity of the band structure, the low-energy electronic properties of TMDs in the vicinity of the two valleys $\pm K$ may be understood within a minimal band model \cite{Xiao, Yuan}. Based on general symmetry consideration, the conduction bands of MoS$_2$ near the valleys are mainly dominated by the Mo $4d_{z^2}$ orbitals \cite{Cappelluti}. In the basis of ($c_{k \uparrow}$, $c_{k \downarrow}$), the two-band $k \cdot p$ Hamiltonian of TMDs material near $K$ and $-K$ valleys can be approximately expressed as \cite{Law1, Yuan}:
\begin{eqnarray}
&&  H_{\pm} = \frac{\hbar^2 k^2}{2m} - \mu \pm \beta \sigma_z.
\end{eqnarray}
Here, $c_{k \uparrow / \downarrow}$ is the annihilation operator of $4d_{z^2}$ electrons, $\pm$ is for the $K$ and $-K$ valleys, and $\sigma_z$ is the Pauli matrix of the spin space. $\mu$ is the chemical potential and $\beta$ is the Ising SOC strength. The conduction band minima appear at two valleys. Ising SOC has opposite directions in opposite valleys and preserves time-reversal symmetry [see Fig. 1(c)], which would induce spin-triplet pairing correlations in the in-plane directions \cite{Law1}. This model for the conduction band can also be applied to other TMDs. For the considered one-dimensional F/ISC junction and F/ISC/F junction, we set the chemical potential $\mu=\mu_S$ in the ISC region and $\mu=\mu_F$ in the F region. Note that the potential at the interface of the junction is neglected in the following calculation, which may have slight effect on the numerical results (see the appendix).

\begin{figure}
\includegraphics[width=8.0cm,height=7.0cm]{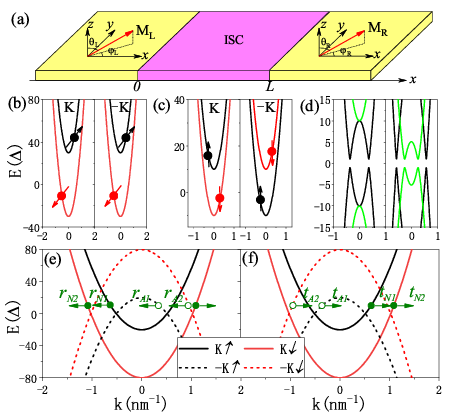}
\caption{ (a) The schematic diagram for the F/ISC/F junction. (b) The bands for different spins in the F region with $\mu_F=0$. (c) The bands for the TMDs in the normal phase with $\mu_S=0$. (d) The bands for the Ising superconducting phase. Both single band ($\mu_S=0$, left) and double band ($\mu_S=15$, right) are exhibited. (e) and (f) The bands for (e) LAR and (f) CAR at $\mu_F=50$. The black curves are for the spin subbands parallel to the magnetization vector $\mathbf{M}$, while the red curves are for the spin subbands antiparallel to $\mathbf{M}$ at both valleys. Other parameters are $M=30$ and $\beta=10$.}
\end{figure}

The Bogoliubov-de Gennes (BdG) Hamiltonians for the ISC region can be written as:
\begin{eqnarray}
H_{\pm}^S (k) = \left(\begin{array}{cc}  H_{\pm}(k)  &  \hat{\Delta} (k)  \\  -\hat{\Delta}^{\ast} (-k)  &  -H_{\mp}^{\ast} (-k) \end{array}\right),
\end{eqnarray}
where $\hat{\Delta} (k)=\Delta e^{i \phi} i \sigma_y$ is the superconducting order parameter with the superconducting gap $\Delta$. The energy band near the valleys of ISC in the normal phase is shown in Fig. 1(c) and the bands for different spins are opposite at opposite valleys. Fig. 1(d) presents the energy bands for the Ising superconducting phase. The ISC is double-band when $\mu_S > \beta$ and is single-band when $\mu_S < \beta$ which can be controlled by the gate voltage in experiment. The Ising superconducting phase would greatly effect the AR process in the F/ISC/F junction. The dispersion relation of the ISC $H_{+}^S (k)$ can be described as:
\begin{eqnarray}
E^2= \Delta^2 + (\frac{\hbar^2 k^2}{2m} - \mu_S \pm \beta)^2,
\end{eqnarray}
where $\pm$ are for an electron with spin up and spin down, respectively. The solutions for $H_{+}^S (k)$ are:
\begin{eqnarray}
g_1 e^{\pm i k_{e1} x}, \hspace{2mm} g_2 e^{\pm i k_{e2} x}, \hspace{2mm} g_3 e^{\pm i k_{h1} x}, \hspace{2mm} g_4 e^{\pm i k_{h2} x}.
\end{eqnarray}
Here, $g_1=(ue^{i\phi/2}, 0, 0, ve^{-i\phi/2})^T$, $g_2=(0, -ue^{i\phi/2}, ve^{-i\phi/2}, 0)^T$, $g_3=(ve^{i\phi/2}, 0, 0, ue^{-i\phi/2})^T$, and $g_4=(0, -ve^{i\phi/2}, ue^{-i\phi/2}, 0)^T$. $u=\sqrt{(E+\Omega)/2E}$, $v=\sqrt{(E-\Omega)/2E}$, $\Omega=\sqrt{E^2-\Delta^2}$, $k_{e1(2)}=\sqrt{2m(\Omega+\mu_S-(+)\beta)}/\hbar$, and $k_{h1(2)}=\sqrt{2m(-\Omega+\mu_S-(+)\beta)}/\hbar$. For the solutions of $H_{-}^S (k)$, the wave vectors $k_{e1(2)}$ and $k_{h1(2)}$ should become $k_{e1(2)}=\sqrt{2m(\Omega+\mu_S+(-)\beta)}/\hbar$ and $k_{h1(2)}=\sqrt{2m(-\Omega+\mu_S+(-)\beta)}/\hbar$.

The BdG Hamiltonians for the F region read:
\begin{eqnarray}
H_{\pm}^F (k) = \left(\begin{array}{cc}  H_{\pm}(k)+\sigma\cdot\mathbf{M}_{L,R}  &  0  \\  0  &  -H_{\mp}^{\ast} (-k) - \sigma^{\ast}\cdot\mathbf{M}_{L,R} \end{array}\right),\nonumber\\
\end{eqnarray}
where $\mathbf{M}_{L,R}=M_{L,R}(\sin\theta_{L,R}\cos\varphi_{L,R}, \sin\theta_{L,R}\sin\varphi_{L,R},$ $\cos\theta_{L,R})$ is the magnetization vector with the exchange energy $M_{L,R}$, the polar angle $\theta_{L,R}$, and the azimuthal angle $\varphi_{L,R}$, which would break the time-reversal symmetry. $\sigma=(\sigma_x, \sigma_y, \sigma_z)$ is the Pauli matrix for spin index. The Ising SOC strength $\beta$ is assumed to be zero in the F region. The magnetization direction can be continuously changed by a small magnetic field. The following results prove that the spin-triplet AR is strongly dependent on the magnetization direction. Opposite to the Ising SOC, the magnetization could lift the spin degeneracy in the subbands and the spin splitting has the same direction at both valleys [see Fig. 1(b)]. The dispersion relation of the F region can be described as:
\begin{eqnarray}
E_{e(h)}= +(-)\frac{\hbar^2 q^2}{2m} -(+) \mu_F \pm M,
\end{eqnarray}
where $\pm$ represent an electron with spin parallel and antiparallel magnetization, respectively. Given the Fermi energy $E$, the eigenvectors of BdG Hamiltonian describing electrons and holes in the F region can be obtained:
\begin{eqnarray}
f_1 e^{\pm i q_{e1} x}, \hspace{2mm} f_2 e^{\pm i q_{e2} x}, \hspace{2mm} f_3 e^{\pm i q_{h1} x}, \hspace{2mm} f_4 e^{\pm i q_{h2} x}.
\end{eqnarray}
Here, $f_1=(\alpha_1, \alpha_2, 0, 0)^T$, $f_2=(-\alpha_2^{\ast}, \alpha_1, 0, 0)^T$, $f_3=(0, 0, \alpha_1, \alpha_2^{\ast})^T$, and $f_4=(0, 0, -\alpha_2, \alpha_1)^T$. $\alpha_1=\cos(\theta_L/2)$, $\alpha_2=\sin(\theta_L/2)e^{i\varphi_L}$, $q_{e1(2)}=\sqrt{2m(E+\mu_L-(+)M_L)}/\hbar$, and $q_{h1(2)}=\sqrt{-2m(E-\mu_L+(-)M_L)}/\hbar$. The eigenvalues and eigenvectors of $H_{+}^F (k)$ are the same as the ones of $H_{-}^F (k)$.

Considering an incident electron from $K$ valley subjected to antiparallel magnetization, the wave functions in the F/ISC junction can be written as
\begin{eqnarray}
\psi_{Fl} &=& f_2 e^{i q_{e2} x} + r_{N1} f_1 e^{-i q_{e1} x} + r_{N2} f_2 e^{-i q_{e2} x} \nonumber\\
&& + r_{A1} f_3 e^{i q_{h1} x} + r_{A2} f_4 e^{i q_{h2} x}, \\
\psi_S &=& t_1 g_1 e^{i k_{e1} x} + t_2 g_2 e^{i k_{e2} x} + t_3 g_3 e^{-i k_{h1} x} + t_4 g_4 e^{-i k_{h2} x}. \nonumber
\end{eqnarray}
Here, $r_{N1}$, $r_{N2}$, $r_{A1}$, and $r_{A2}$ are the various reflection coefficients. Based on the boundary conditions $\psi_{Fl}(x=0)=\psi_S(x=0)$ and $\psi'_{Fl}(x=0)=\psi'_S(x=0)$, we can obtain four reflectivities:
\begin{eqnarray}
&& R_{N1}=Re(\frac{q_{e1}}{q_{e2}})|r_{N1}|^2, \hspace{2mm} R_{N2}=|r_{N2}|^2, \nonumber\\
&& R_{A1}=Re(\frac{q_{h1}}{q_{e2}})|r_{A1}|^2, \hspace{2mm} R_{A2}=Re(\frac{q_{h2}}{q_{e2}})|r_{A2}|^2.
\end{eqnarray}

The wave functions in the F/ISC/F junction can be written as
\begin{eqnarray}
\psi_{Fl} &=& f_2 e^{i q_{e2} x} + r_{N1} f_1 e^{-i q_{e1} x} + r_{N2} f_2 e^{-i q_{e2} x} \nonumber\\
&& + r_{A1} f_3 e^{i q_{h1} x} + r_{A2} f_4 e^{i q_{h2} x}, \nonumber\\
\psi_S &=& t_1 g_1 e^{i k_{e1} x} + t_2 g_2 e^{i k_{e2} x} + t_3 g_3 e^{-i k_{h1} x} + t_4 g_4 e^{-i k_{h2} x} \nonumber\\
&& + t_5 g_1 e^{-i k_{e1} x} + t_6 g_2 e^{-i k_{e2} x} + t_7 g_3 e^{i k_{h1} x} + t_8 g_4 e^{i k_{h2} x}. \nonumber\\
\psi_{Fr} &=& t_{N1} f_5 e^{i q_{e3} x} + t_{N2} f_6 e^{i q_{e4} x} + t_{A1} f_7 e^{-i q_{h3} x}  \nonumber\\
&&  + t_{A2} f_8 e^{-i q_{h4} x}.
\end{eqnarray}
Here, $f_5=(\alpha_3, \alpha_4, 0, 0)^T$, $f_6=(-\alpha_4^{\ast}, \alpha_3, 0, 0)^T$, $f_7=(0, 0, \alpha_3, \alpha_4^{\ast})^T$, and $f_8=(0, 0, -\alpha_4, \alpha_3)^T$. $\alpha_3=\cos(\theta_R/2)$, $\alpha_4=\sin(\theta_R/2)e^{i\varphi_R}$, $q_{e3(4)}=\sqrt{2m(E+\mu_R-(+)M_R)}/\hbar$, and $q_{h3(4)}=\sqrt{-2m(E-\mu_R+(-)M_R)}/\hbar$. Based on the boundary conditions $\psi_{Fl}(x=0)=\psi_S(x=0)$, $\psi_S(x=L)=\psi_{Fr}(x=L)$, $\psi'_{Fl}(x=0)=\psi'_S(x=0)$, and $\psi'_S(x=L)=\psi'_{Fr}(x=L)$, we can obtain the reflectivity and the transmittivity. The functions $\psi'_{Fl}$, $\psi'_S$ and $\psi'_{Fr}$ are the derivatives of $\psi_{Fl}$, $\psi_S$ and $\psi_{Fr}$ with respect to $x$, respectively. The four transmittivities are defined as:
\begin{eqnarray}
&& T_{N1}=Re(\frac{q_{e3}}{q_{e2}})|t_{N1}|^2, \hspace{2mm} T_{N2}=Re(\frac{q_{e4}}{q_{e2}})|t_{N2}|^2, \nonumber\\
&& T_{A1}=Re(\frac{q_{h3}}{q_{e2}})|t_{A1}|^2, \hspace{2mm} T_{A2}=Re(\frac{q_{h4}}{q_{e2}})|t_{A2}|^2.
\end{eqnarray}

When an electron with spin antiparallel magnetization from $K$ valley moves to the ISC interface, as shown in Figs. 1(e) and 1(f), $R_{N1}$ is the anomalous electron reflection due to spin flip, $R_{N2}$ is the normal electron reflection, $R_{A1}$ is the normal AR, and $R_{A2}$ is the equal-spin AR due to the equal-spin-triplet pairing where the incident electron and the reflected hole belong to the same spin. $T_{N1}$ and $T_{N2}$ are anomalous and normal electron transport, respectively. $T_{A1}$ and $T_{A2}$ are normal and anomalous CAR, respectively. For incident electron with other spin from other valley, the reflectivity and transmittivity for electron and hole can be similarly obtained by the BdG equation and boundary condition. Subsequently, for electrons with spin parallel and antiparallel the magnetization directions near both valleys, the total LAR is defined as:
\begin{eqnarray}
R_A= \sum_{K,-K}\sum_{M,-M} ( R_{A1}+R_{A2} ).
\end{eqnarray}
The total CAR is defined as:
\begin{eqnarray}
T_A= \sum_{K,-K}\sum_{M,-M} ( T_{A1}+T_{A2} ).
\end{eqnarray}
The summation is to sum all the electrons with both spins near both valleys $\pm K$. The two spins are parallel and antiparallel to the magnetization, labeled by $M$ and $-M$, respectively.

\section{Results and discussions}
In the following calculation, the superconducting gap is fixed as $\Delta_S=1.0meV$. $\Delta_S$ is the unit of $M_{L,R}$, $\beta$, $\mu_S$, and $\mu_{R,L}$. The unit of width $L$ is the superconducting coherence length $\xi=\hbar v_F/\pi \Delta$. For convenience, the parameters are set as $\mu_L=\mu_R=\mu_F$, $M_L=M_R=M=30$, $\beta=10$, and $\varphi_L=\varphi_R=0$. The incident energy is $E=0$ and the width is $L=0.5\xi$, unless otherwise noted. Note that we consider an idealized junction, where the mass is assumed to be the same in ISC and F, and F has no SOC. The different masses would effect momentum-energy relation and the SOC in F would lock the spin and valley degrees of freedom. However, this should have no substantial effect on the discussed equal-spin CAR, oblique-spin CAR, and the main conclusion, which depend mainly on the half-metal F, the magnetization direction, and the Ising superconducting phase (see the appendix). We will first discuss the equal-spin LAR in F/ISC junction in subsection III.A, then discuss the equal-spin CAR in F/ISC/F junction with $\theta_L=\theta_R$ in subsection III.B and the oblique-spin CAR in F/ISC/F junction with $\theta_L \neq \theta_R$ in subsection III.C.

\subsection{Equal-spin LAR in F/ISC junction}

\begin{figure}
\includegraphics[width=8.0cm,height=6.0cm]{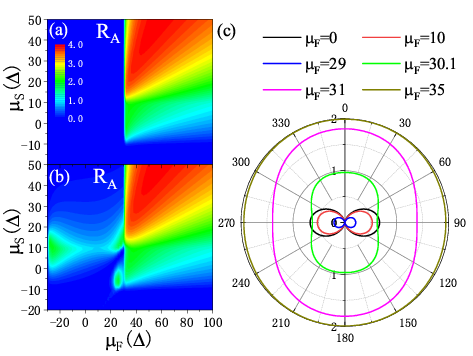}
\caption{ LAR coefficient $R_A$ versus potentials $\mu_F$ and $\mu_S$ in F/ISC junction when (a) $\theta_L=0$ and (b) $\theta_L=\pi/2$. (c) $R_A$ versus $\theta_L$ when $\mu_S=5$.}
\end{figure}

Firstly, we discuss the LAR in F/ISC junction. The contour plot for LAR coefficient $R_A(\mu_F, \mu_S)$ is shown in Figs. 2(a) and 2(b). It is evident that when the magnetization vector $\mathbf{M}$ is perpendicular to the junction plane with $\theta_L=0$, there is no LAR in the region $\mu_F<M$, since the F is completely spin polarized and the Cooper pair for AR requires the electrons from opposite spins. As $\mu_F$ increases and $\mu_F>M$, the spin polarization of F reduce to zero, where both spin-up and spin-down subbands could exist, thus the normal LAR takes place. The ISC is single-band at $-\beta<\mu_S<\beta$, and there are only two channels contributed by $K\downarrow$ and $-K\uparrow$ states, leading to the LAR coefficient $R_A\approx2$. The ISC becomes double-band at $\mu_S>\beta$ and four channels are opened by $K\downarrow$, $-K\uparrow$, $K\uparrow$, and $-K\downarrow$ states, thus $R_A\approx4$ [see Fig. 2(a)].

However, when the magnetization direction is parallel to the junction plane with $\theta_L=\pi/2$, dramatically, the LAR coefficient $R_A$ is finite in the spin polarized F with $\mu_F<M$ due to the appearance of equal-spin-triplet AR [see Fig. 2(b)], which consists with the one of tight-binding model \cite{Sun3}. Because of Ising SOC, the spin-singlet and spin-triplet Cooper pairs can be generated in ISC \cite{Law1}. The equal-spin-triplet Cooper pairs $| \rightrightarrows \rangle$ and $| \leftleftarrows \rangle$ have electron spins pointing to the in-plane directions. When an incident electron $| \rightarrow \rangle$ with spin polarized in the in-plane directions from one valley is injected into the ISC, it could compose equal-spin-triplet Cooper pairs $| \rightrightarrows \rangle$ by the same spins but opposite valleys in ISC. Simultaneously, a hole state $| \rightarrow \rangle$ is formed and reflected in the F with the spin pointing to in-plane direction, resulting in the equal-spin-triplet AR. Distinctly, this anomalous AR mainly occurs in the spin polarized region $\mu_F<M$.

Fig. 2(c) shows the dependence of LAR on the magnetization direction. For half-metal F with $\mu_F<M$, $R_A$ is devoted by the equal-spin AR and the normal AR process is completely suppressed, thus, $R_A=0$ at $\theta_L=0, \pi$. On the contrary, $R_A$ reaches its maximum at $\theta_L=\pi/2, 3\pi/2$ because the Cooper pairs have the spin-triplet components $| \leftleftarrows \rangle$ and $| \rightrightarrows \rangle$ with spin pointing to in-plane directions. The magnetoanisotropy of $R_A$ is very prominent for half-metal F, the period of which is $\pi$. Oppositely, $R_A$ is mainly managed by the normal AR at $\mu_F>M$, and so $R_A$ presents its maximum at $\theta_L=0, \pi$. As $\mu_F$ increases, $R_A$ trends to magnetoisotropy.

\subsection{Equal-spin CAR in F/ISC/F junction with $\theta_L=\theta_R$}

\begin{figure}
\includegraphics[width=8.0cm,height=7.0cm]{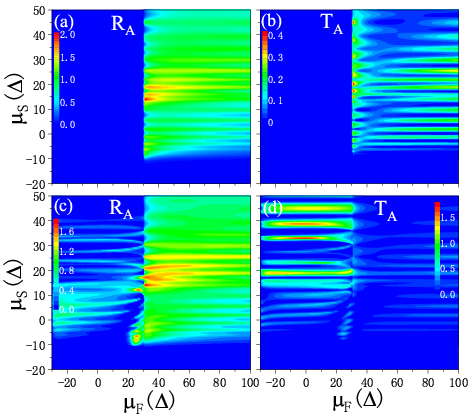}
\caption{ LAR coefficient $R_A$ and CAR coefficient $T_A$ versus potentials $\mu_F$ and $\mu_S$. $\theta=0$ in (a,b) and $\theta=\pi/2$ in (c,d).}
\end{figure}

Now, we turn to study the properties of equal-spin CAR in F/ISC/F junction and the results are shown in Figs. 3-7. We set $\theta_L=\theta_R=\theta$ in this subsection, that is, the magnetization directions of the left and right F leads are the same. Considering the Meissner effect of SC, the magnetization direction of the left and right F leads can be simultaneously adjusted by the magnetic field in experiment, and the ISC is not affected. Below it is demonstrated that CAR can be controlled by changing the magnetization direction.

Fig. 3 displays the contour plot for LAR coefficient $R_A$ and CAR coefficient $T_A$ as a function of the potentials $\mu_F$ and $\mu_S$ when (a)(b) $\theta=0$ and (c)(d) $\pi/2$. One may find that $R_A$ is zero at $\theta=0$ for half-metal Fs with $\mu_F<M$ [see Fig. 3(a)]. Furthermore, Fig. 3(b) shows that the normal CAR coefficient is completely suppressed, that is $T_A=0$ at $\theta=0$ and $\mu_F<M$, because both left and right F leads are spin polarized half-metals with the same spin polarization, but the Cooper pair requires opposite spins in $z$ direction from the two leads.
As expected, the normal LAR and CAR could appear when $\mu_F>M$. Interestingly, both LAR and CAR show oscillatory behavior with $\mu_S$ due to quantum interference effects of the electronic states in ISC, different from the result of F/ISC junction. More resonance peaks could be formed with the increase of the width $L$. When the magnetization direction in the Fs is parallel to the junction plane with $\theta=\pi/2$, it is expected that the equal-spin LAR could occur. Significantly, an interesting equal-spin CAR also appears in the region $\mu_F<M$ [see Figs. 3(c) and 3(d)]. Different from the equal-spin LAR in F/ISC junction which mainly occurs in the single-band region, the equal-spin LAR and equal-spin CAR in F/ISC/F junction appear in both single-band and double-band regions arising from quantum interference. In particular, the equal-spin CAR is rather small for single-band ISC but very large for double-band ISC. Oppositely, the equal-spin LAR is rather large for single-band ISC but greatly weakened for double-band ISC. Furthermore, compared with the normal CAR at $\theta=0$, one can clearly see that the equal-spin CAR at $\theta=\pi/2$ is very strong. However, the equal-spin LAR is quiet weak compared with the normal LAR. The physical mechanism for equal-spin LAR in the F/ISC/F junction is similar to that in F/ISC junction. The equal-spin CAR can be understood as follows. When the incident electron $| \rightarrow \rangle$ with the spin pointing to in-plane direction injects into ISC from the left F lead, it could compose equal-spin-triplet Cooper pair $| \rightrightarrows \rangle$ with the electron $| \rightarrow \rangle$ with the same spin from the right F lead due to the Ising SOC. Meanwhile, a hole state $| \rightarrow \rangle$ is transmitted in the right F lead, leading to the equal-spin CAR.

\begin{figure}
\includegraphics[width=8.0cm,height=9.0cm]{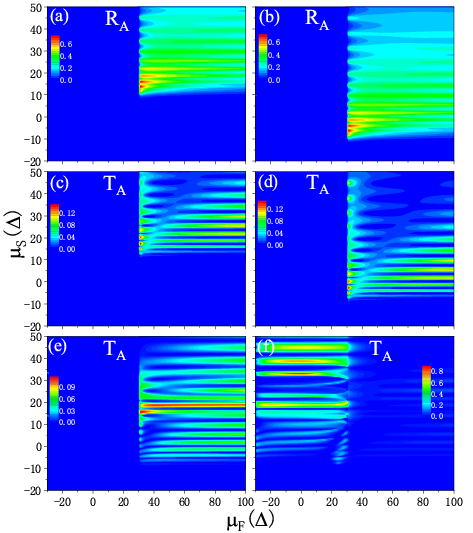}
\caption{ (a-d) LAR coefficient $R_A$ and CAR coefficient $T_A$ versus $\mu_F$ and $\mu_S$ by (a,c) $K \uparrow$ and (b,d) $K \downarrow$ electrons with $\theta=0$. (e) and (f) CAR coefficient $T_A$ versus $\mu_F$ and $\mu_S$ by (e) $K \rightarrow$ and (f) $K \leftarrow$ electrons with $\theta=\pi/2$.}
\end{figure}

In order to understand deeply the effect of spin polarized F and Ising superconducting phase on the AR, we study spin- and valley-dependent LAR and CAR. Fig. 4 shows the normal LAR and normal CAR contributed by (a)(c) $K \uparrow$ and (b)(d) $K \downarrow$ electrons when $\theta=0$. Thanks to the time-reversal symmetry in the ISC, the state $|K \uparrow \rangle$ is degenerate with $|-K \downarrow \rangle$. The states $|K \downarrow \rangle$ and $|-K \uparrow \rangle$ are similarly degenerate. As a consequence, $R_A(K \uparrow)=R_A(-K \downarrow)$, $R_A(K \downarrow)=R_A(-K \uparrow)$, $T_A(K \uparrow)=T_A(-K \downarrow)$, and $T_A(K \downarrow)=T_A(-K \uparrow)$. The total $R_A$ and $T_A$ by the two spins at two valleys are the results in Figs. 3(a) and 3(b), respectively. Obviously, the normal LAR and normal CAR occur only in the region $\mu_F>M$. When the Ising superconducting phase is single-band with $\mu_S<\beta$, only $K \downarrow$ and $-K \uparrow$ incident electrons can form Cooper pairs $| \uparrow \downarrow \rangle$ since the single-band comes from the $K \downarrow$ and $-K \uparrow$ states [see Fig. 1(c)]. As a result, the normal LAR and CAR in the single-band region are mainly contributed by the $K \downarrow$ and $-K \uparrow$ states [see Figs. 4(b) and 4(d)]. The normal LAR and CAR by $K \uparrow$ and $-K \downarrow$ just occur in the double-band region $\mu_S>\beta$ [see Figs. 4(a) and 4(c)]. Compared with LAR, the oscillation behavior of CAR is stronger.

Figs. 4(e) and 4(f) show the equal-spin CAR contributed by $K \rightarrow$ and $K \leftarrow$ electrons when $\theta=\pi/2$, respectively. All the incident electrons $| \rightarrow \rangle$ and $| \leftarrow \rangle$ could form the equal-spin-triplet Cooper pairs and the equal-spin CAR could occur in both singe-band and double-band regions when $\theta=\pi/2$. In particular, the equal-spin LAR and CAR are independent of the valley degree of freedom, $T_A(K \rightarrow)=T_A(-K \rightarrow)$, $T_A(K \leftarrow)=T_A(-K \leftarrow)$, $R_A(K \rightarrow)=R_A(-K \rightarrow)$, and $R_A(K \leftarrow)=R_A(-K \leftarrow)$ (not shown in the figure), oppositing to the case of normal LAR and CAR when $\theta=0$. When spin is fully polarized in F leads with $\mu_F<M$, the state $| \rightarrow \rangle$ is evanescent state and so $T_A(K \rightarrow)=T_A(-K \rightarrow)=0$ [see Fig. 4(e)]. However, the equal-spin CAR with a large coefficient by $| \leftarrow \rangle$ mainly appears in the region $\mu_F<M$ responsible for the propagating state $| \leftarrow \rangle$ in the F region [see Fig. 4(f)]. Obviously, the equal-spin CAR is dramatically enhanced in the half-metal F.

\begin{figure}
\includegraphics[width=8.0cm,height=6.0cm]{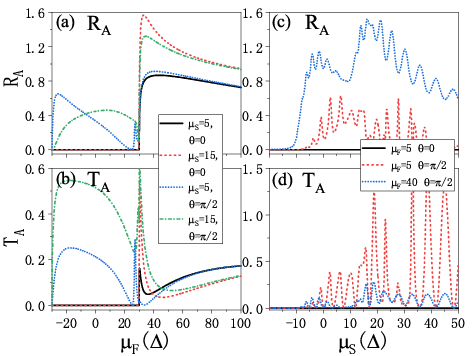}
\caption{ LAR coefficient $R_A$ and CAR coefficient $T_A$ versus (a,b) $\mu_F$ and (c,d) $\mu_S$. }
\end{figure}

Fig. 5 presents the LAR coefficient $R_A$ and CAR coefficient $T_A$ as a function of (a)(b) $\mu_F$ and (c)(d) $\mu_S$. Figs. 5(a) and 5(b) show that $R_A=T_A=0$ at $\theta=0$ for half-metal Fs with $\mu_F<M$. However, $R_A$ and $T_A$ have considerable values at $\theta=\pi/2$ and $\mu_F<M$ due to the equal-spin LAR and equal-spin CAR. When $\mu_F>M$, $R_A$ and $T_A$ are insensitive to the change of magnetization angle $\theta$ owing to the normal LAR and normal CAR. Furthermore, the CAR coefficient exhibits a remarkable peak near $\mu_F=M$, because that the density of states of the F leads are singular at $\mu_F=M$. Different from the ones in Figs. 5(a) and 5(b), LAR and CAR oscillate with $\mu_S$, as shown in Figs. 5(c) and 5(d). For half-metal F with $\mu_F=5$ when $\theta=\pi/2$, the CAR is very strong with a large amplitude but the LAR is relatively small. Oppositely, the CAR become weak while the LAR become strong when $\mu_F>M$. In addition, LAR and CAR present an obvious step behavior near the boundary of single-band and double-band Ising superconductivity when $\mu_F>M$, since more channels are opened by the electronic states [see the dotted curves in Figs. 5(c) and 5(d)].

\begin{figure}
\includegraphics[width=8.0cm,height=4.0cm]{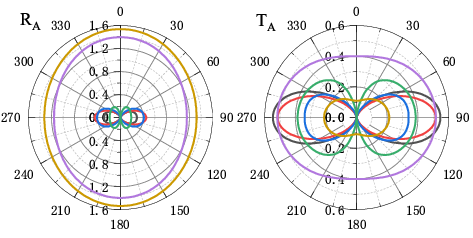}
\caption{ LAR coefficient $R_A$ and CAR coefficient $T_A$ versus the magnetization direction $\theta$ when $\mu_S=15$. The black, red, blue, green, purple, yellow curves are for $\mu_F=-20$, $0$, $20$, $29$, $31$, and $35$. }
\end{figure}

Fig. 6 discusses the magnetoanisotropy phenomenon of the LAR and the CAR, the change of which are not monotonic with $\theta$. From Fig. 6(b) we can clearly see that the CAR process displays a remarkable magnetoanisotropy and the period is $\pi$. The CAR coefficient $T_A$ is symmetric about $\theta=0, \pi/2$, satisfying $T_A(\theta)=T_A(\pi\pm\theta)=T_A(2\pi-\theta)$. For half-metal F with $\mu_F<M$, $T_A$ only comes from the contribution of equal-spin CAR. When the magnetization direction in the Fs deviates from $\theta=0$, the Ising SOC will play a part in the spin-flip scattering at the interface, leading to the appearance of the equal-spin CAR. $T_A=0$ at $\theta=0$ since the equal-spin CAR is absent. By rotating the magnetization direction $\theta$ from $\theta=0$, $T_A$ first increases, achieving its maximum at $\theta=\pi/2$ due to the strong spin-flip, and then decreases. Dramatically, the magnetoanisotropic equal-spin CAR can be switched on and off by changing the magnetization direction in the half-metal limit. When $\mu_F>M$, the normal CAR happens and CAR coefficient $T_A$ at $\theta=0$ is no longer zero. Thus, the CAR process would trend to magnetoisotropy. The magnetoanisotropy property of LAR process is analogous to the CAR process [see Fig. 6(a)]. In addition, the position of maximum for LAR changes to $\theta=0$ when $\mu_F>M$, implying a strong normal LAR. The position of maximum for CAR is always $\theta=\pi/2$, suggesting a dominant role of equal-spin CAR.

\begin{figure}
\includegraphics[width=8.0cm,height=7.0cm]{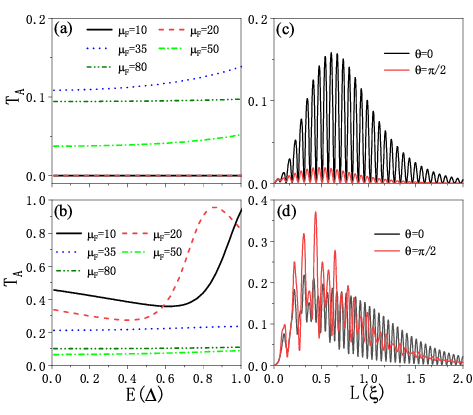}
\caption{ CAR coefficient $T_A$ versus incident energy $E$ when (a) $\theta=0$ and (b) $\theta=\pi/2$ with $\mu_S=15$ and $L=0.5\xi$. CAR coefficient $T_A$ versus width $L$ when (c) $\mu_S=5$ and (d) $\mu_S=15$ with $E=0$ and $\mu_F=35$. }
\end{figure}

At the end of this subsection, we discuss the effect of Fermi energy $E$ and width $L$ on the CAR in Fig. 7, which is important for the experimental observation. When $\theta=0$ in Fig. 7(a), CAR as a function of superconducting gap energy is $T_A=0$ at $\mu_F<M$ because the spin is conserved in the $z$ direction and the equal-spin CAR cannot take place. The CAR is finite at $\mu_F>M$ due to the occurrence of normal CAR. When $\theta=\pi/2$ in Fig. 7(b), the equal-spin CAR could occur and so $T_A$ has a large value at $\mu_F<M$. In addition, with the increase of the potential $\mu_F$, CAR becomes insensitive to both magnetization angle and Fermi energy. The CAR as a function of width $L$ at $\mu_S=5$ and $\mu_S=15$ is displayed in Figs. 7(c) and 7(d), respectively. The results indicate that the CAR is sensitive to the width $L$. At $\mu_S=5$ and $\theta=0$, only normal CAR could occur which is devoted by $K\downarrow$ and $-K\uparrow$ electrons. Thus, CAR exhibits a regular oscillation with large amplitude as the width $L$ increases [see the black curve in Fig. 7(c)]. The resonant peaks of CAR arise from the quantum interference of the states in ISC region. At $\theta=\pi/2$, interesting, the CAR has the same oscillation period with the one at $\theta=0$ but its amplitude is small. For the double-band situation with $\mu_S=15$ in Fig. 7(d), the CAR is contributed by more channels with different periods and different amplitudes, and so it presents an irregular oscillation.

\subsection{Oblique-spin CAR in F/ISC/F junction with $\theta_L \neq \theta_R$}

\begin{figure}
\includegraphics[width=8.0cm,height=4.0cm]{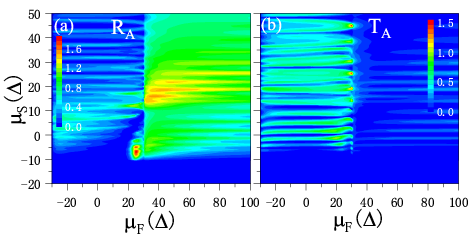}
\caption{ LAR coefficient $R_A$ and CAR coefficient $T_A$ versus potentials $\mu_F$ and $\mu_S$ with $\theta_L=\pi/2$ and $\theta_R=0$. }
\end{figure}

When the magnetization directions in the left and right F leads are different, the LAR process and CAR process will become more interesting, as shown in Figs. 8-12. When $\theta_L=\pi/2$ and $\theta_R=0$ in Fig. 8, the remarkable LAR and CAR could arise at $\mu_F<M$, and the CAR process is very strong. Compared with the ones observed in Figs. 3(c) and 3(d), more resonance peaks are formed in LAR and CAR. Note that the CAR process in Fig. 8(b) is no longer equal-spin CAR while the LAR process is still equal-spin LAR. Here the incident electron with spin $| \leftarrow \rangle$ from the left F lead composes Cooper pair with the spin $|\downarrow\rangle$ electron from the right F lead, and the spin of the reflected hole points to $|\downarrow\rangle$ direction. Thereafter we name the CAR with the spins of the electron and hole neither parallel nor antiparallel as the oblique-spin CAR.

\begin{figure}
\includegraphics[width=8.0cm,height=7.0cm]{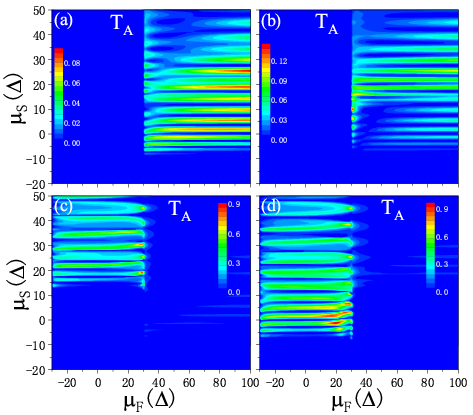}
\caption{ CAR coefficient $T_A$ versus $\mu_F$ and $\mu_S$ by (a) $K\rightarrow$, (b) $-K\rightarrow$, (c) $K\leftarrow$, and (d) $-K\leftarrow$ electrons. The parameters are the same as those in Fig. 8.}
\end{figure}

Fig. 9 exhibits the CAR devoted by (a) $K\rightarrow$, (b) $-K\rightarrow$, (c) $K\leftarrow$, and (d) $-K\leftarrow$ electrons when $\theta_L=\pi/2$ and $\theta_R=0$. The total $T_A$ by the four kinds of electron is the result in Fig. 8(b). Interestingly, the CAR by the two spins near the two valleys occurs in different areas of the ($\mu_F, \mu_S$) space. The CAR coefficient $T_A$ by $K\rightarrow$ and $-K\rightarrow$ only occurs in the region $\mu_F>M$, because they have no electronic states in the half-metal Fs when $\mu_F<M$ [see Figs. 9(a) and 9(b)]. $T_A(-K\rightarrow)$ mainly appears in the double-band region. However, CAR by $K\leftarrow$ and $-K\leftarrow$ could occur in the spin polarized F with $\mu_F<M$. Distinctly, $T_A(-K\leftarrow)$ could appear in the single-band region while $T_A(K\leftarrow)$ mainly appears in the double-band region [see Figs. 9(c) and 9(d)]. In the spin polarized half-metal F at $\theta_L=\pi/2$, the states in the left F lead satisfy $| \rightarrow \rangle = \sqrt{2} (|\uparrow \rangle + |\downarrow \rangle)/2$ and $| \leftarrow \rangle = \sqrt{2} (|\uparrow \rangle - |\downarrow \rangle)/2$, which would split into $|\uparrow \rangle$ and $|\downarrow \rangle$ states at the ISC interface. The incident electrons $K\leftarrow$ and $-K\leftarrow$ with an in-plane spin would split into $K \uparrow,\downarrow$ and $-K \uparrow,\downarrow$ electronic states at the left interface, respectively. When the Ising superconducting phase is single-band at $\mu_S<\beta$, only $K \downarrow$ and $-K \uparrow$ states could tunnel to the right interface. On the other hand, only $K \downarrow$ and $-K \downarrow$ electronic states exist in the right F at $\theta_R=0$ and $\mu_F<M$. Thus, only $-K \uparrow$ state from the left could compose Cooper pair with the $K \downarrow$ state from the right, leading to the CAR by $-K\leftarrow$ electron in the single-band region, as shown in Fig. 9(d). When the Ising superconducting phase change to double-band at $\mu_S>\beta$, $K \uparrow$ state from the left could also tunnel to the right interface and compose Cooper pair with the $-K \downarrow$ state from the right F, giving rise to the CAR by $K\leftarrow$ electron in the double-band region [see Fig. 9(c)]. When the spin polarization for F reduce to zero at $\mu_F>M$, the $K\rightarrow$ and $-K\rightarrow$ states could compose Copper pair and produce CAR effortlessly [see Figs. 9(a) and 9(b)]. Significantly, from Fig. 9 one may find that the area of CAR and the position of its resonance peak in the ($\mu_F$, $\mu_S$) space strongly depend on the spin and valley degrees of freedom. Therefore, the F/ISC/F junction realizes a spin and valley polarized CAR which can be effectively controlled by the potentials $\mu_F$ and $\mu_S$.

\begin{figure}
\includegraphics[width=8.0cm,height=8.0cm]{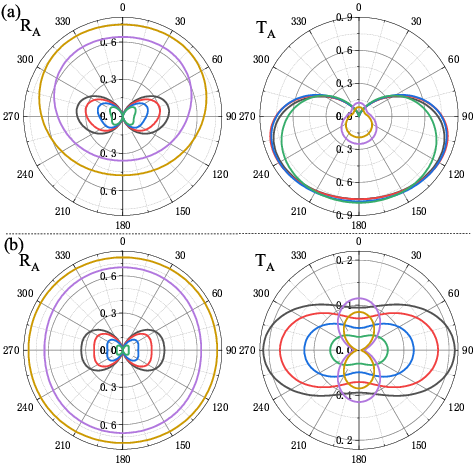}
\caption{ LAR coefficient $R_A$ and CAR coefficient $T_A$ versus the magnetization direction $\theta_L$ with (a) $\theta_R=0$ and (b) $\theta_R=\pi/2$ when $\mu_S=5$. The black, red, blue, green, purple, yellow curves are for $\mu_F=0$, $8$, $15$, $20$, $31$, and $32$. }
\end{figure}

Next, we study the anisotropic behavior of LAR and CAR with respect to the magnetization angles $\theta_L$ and $\theta_R$, as shown in Figs. 10-12. Fig. 10 displays the LAR and CAR as a function of $\theta_L$ at (a) $\theta_R=0$ and (b) $\theta_R=\pi/2$ for different values of $\mu_F$. When $\theta_R=0$ in Fig. 10(a), the LAR process is mainly equal-spin LAR for half-metal F at $\mu_F<M$, so $R_A=0$ at $\theta_L=0, \pi$. The equal-spin CAR disappears and only oblique-spin CAR could take place at $\theta_R=0$ and $\mu_F<M$. So $T_A=0$ at $\theta_L=\theta_R=0$, and $T_A$ increases when the angle $\theta_L$ deviates from the $z$ direction. LAR is relatively weak at $\mu_F<M$ and it can be enhanced at $\mu_F>M$. On the contrary, CAR is quite strong and robust to $\mu_F$ at $\mu_F<M$ but it is weaken at $\mu_F>M$. The magnetoanisotropy of LAR and CAR is very prominent when $\mu_F<M$ [see Fig. 10(a)]. However, the anisotropy is weakened at $\mu_F>M$ due to the appearance of more AR. Generally, the period of magnetoanisotropy becomes $2\pi$ instead of $\pi$ due to the effect of the right F. Both $R_A$ and $T_A$ are symmetric about $\theta_L=0$. In fact, the feature of magnetoanisotropy for AR with respect to $\theta_L$ can be managed by $\theta_R$, as shown in Fig. 10(b) with $\theta_R=\pi/2$.

When $\theta_R=\pi/2$, one may get $R_A(\theta_L)=R_A(\pi\pm\theta_L)=R_A(2\pi-\theta_L)$ and $T_A(\theta_L)=T_A(\pi\pm\theta_L)=T_A(2\pi-\theta_L)$, as shown in Fig. 10(b). LAR and CAR are symmetric with respect to $\theta_L=0, \pi/2$. $R_A$ is still zero at $\theta_L=0,\pi$ when $\mu_F<M$ since it is equal-spin LAR which is mainly depended on the left F. When $\theta_R=\pi/2$, normal CAR, equal-spin CAR, and oblique-spin CAR could occur depending on the magnetization angle $\theta_L$, so $T_A$ as a function of $\theta_L$ is always nonzero. $T_A$ achieves its maximum at $\theta_L=\pi/2, 3\pi/2$ when $\mu_F<M$ implying a strong equal-spin CAR, while its maximum is at $\theta_L=0, \pi$ when $\mu_F>M$ due to the oblique-spin CAR. Furthermore, LAR is almost isotropy with the increase of $\mu_F$ when $\mu_F>M$, similar to the results in F/ISC/F junction with $\theta_L=\theta_R$.

\begin{figure}
\includegraphics[width=8.0cm,height=7.5cm]{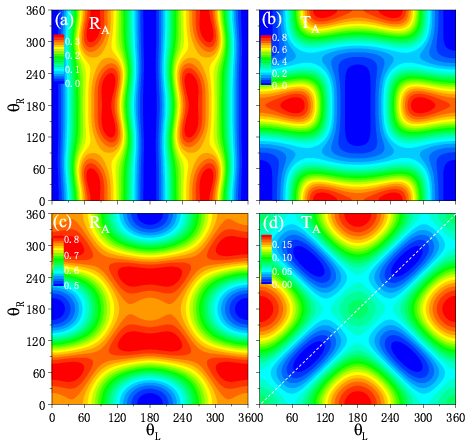}
\caption{ LAR coefficient $R_A$ and CAR coefficient $T_A$ versus the magnetization directions $\theta_L$ and $\theta_R$ with $\mu_S=5$. $\mu_F=5$ in (a,b) and $\mu_F=32$ in (c,d). }
\end{figure}

\begin{figure}
\includegraphics[width=8.0cm,height=7.5cm]{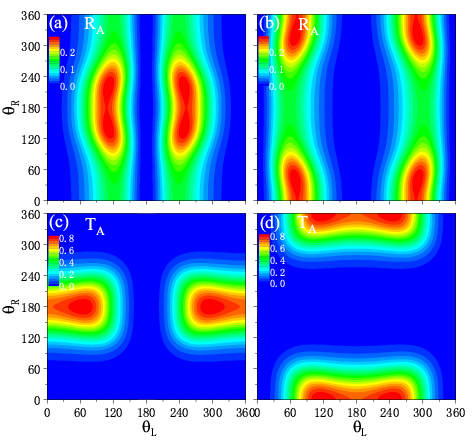}
\caption{ LAR coefficient $R_A$ and CAR coefficient $T_A$ versus the magnetization directions $\theta_L$ and $\theta_R$ by (a,c) $K$ and (b,d) $-K$ electrons. The parameters are the same as those in Figs. 11(a) and 11(b). }
\end{figure}

Fig. 11 presents the LAR and CAR as a function of $\theta_L$ and $\theta_R$ when the Ising superconducting phase is single-band with $\mu_S=5$. As shown in Figs. 11(a) and 11(b) for the half-metal Fs at $\mu_F=5$, the LAR and CAR are contributed by the electron with spin antiparallel magnetization from the two valleys. The equal-spin AR plays a major role in LAR and CAR for the half-metal, the magnetoanisotropy of which is very remarkable. Both LAR and CAR are symmetric about $\theta_{L,R}=0, \pi$. Thus, we have $R_A(\theta_L, \theta_R) = R_A(-\theta_L, -\theta_R) = R_A(\pi+\theta_L, \pi+\theta_R) = R_A(\pi-\theta_L, \pi-\theta_R)$ and $T_A(\theta_L, \theta_R) = T_A(-\theta_L, -\theta_R) = T_A(\pi+\theta_L, \pi+\theta_R) = T_A(\pi-\theta_L, \pi-\theta_R)$. $R_A$ and $T_A$ can achieve extreme values at appropriate angles. For instance, the LAR is $R_A=0$ at $\theta_L=0$, $\pi$ because the equal-spin LAR can not occur at $\theta_L=0$, $\pi$, which is independent of $\theta_R$. The CAR is $T_A=0$ at $\theta_L=\theta_R=0$, $\pi$ due to the absence of normal CAR and equal-spin CAR. In Figs. 11(c) and 11(d) at $\mu_F=32$, the electrons of both spins near both valleys contribute to the AR. The normal LAR could occur and so the minimum is no longer zero in the ($\theta_L$ , $\theta_R$) space [see Fig. 11(c)]. LAR and CAR are still symmetric with respect to $\theta_{L,R}=0, \pi$. According to the spatial symmetry of the F/ISC/F junction and the physical mechanism of CAR, we can find that the left and right magnetization angles $\theta_{L,R}$ have the same effect on CAR, that is $T_A(\theta_L, \theta_R)=T_A(\theta_R, \theta_L)$ [see Fig. 11(d)]. Note that this character still remains in the double-band situation.

The valley dependence of LAR and CAR is discussed in Fig. 12 when $\mu_S=5$ and $\mu_F=5$, which is devoted by the electron with spin antiparallel magnetization from (a)(c) $K$ valley and (b)(d) $-K$ valley. The total $T_A$ (or $R_A$) by $K$ and $-K$ valleys in Fig. 12 is the result in Fig. 11(b) [or Fig. 11(a)]. Significantly, one may clearly see that the CAR contributed by the $K$ and $-K$ valleys appears in different areas of the ($\theta_R, \theta_L$) plane [see Figs. 12(c) and 12(d)]. The LAR has similar results [see Figs. 12(a) and 12(b)]. Therefore, we can achieve the valley polarized CAR and LAR by adjusting the magnetization directions.

Finally, we discuss the experimental realization of the proposed F/ISC/F junction. In fact, many groups have demonstrated that various F/SC junctions could be fabricated in experiment \cite{Buzdin, Bergeret}. Very recently, unconventional supercurrent phase has been reported experimentally in Ising superconductor Josephson junction that couples NbSe$_2$ Ising Cooper pairs across a magnetic insulator Cr$_2$Ge$_2$Te$_6$ \cite{Idzuchi}. An experimental observation on anisotropic magnetoresistance is also reported in ISC/F/ISC junctions that are made of ISC NbSe$_2$ and ferromagnetic insulator CrBr$_3$ \cite{Kang}. In addition, the exchange energy is set to be $30meV$ in the above discussion which can be generated by the ferromagnetic insulator. According to the experiments on SC/F junction, the exchange energy could reach to $86meV$ which is a quite reasonable value \cite{Blum, Kontos}. Therefore, the proposed F/ISC/F junction could be fabricated experimentally, where the equal-spin and oblique-spin CAR should be observable in the present technology.

\section{Conclusion}

In summary, we theoretically studied the AR in F/ISC/F heterojunction by solving the BdG equations. Because of the Ising SOC in ISC, the equal-spin CAR and equal-spin LAR could occur, which present a magnetoanisotropic behaviour with the period $\pi$. The equal-spin CAR could reach its maximum at the in-plane magnetization and decrease to zero at the out-of-plane magnetization. The property of CAR greatly depends on the Ising superconducting phase and magnetization directions, in contrast to the normal SC junction where the CAR is insensitive to the superconducting phase. Furthermore, the spin and valley polarized oblique-spin CAR can be realized by the chemical potentials and magnetization directions. We expect that these findings may be helpful to generate the spin- or valley-entangled states and detect Ising superconductivity.

\section*{appendix}

In this appendix, we discuss the effect of mass difference and interfacial potential on the AR, taking F/ISC junction as an example. The Hamiltonian near valleys can be expressed as \cite{Milanovic}:
\begin{eqnarray}
&&  H_{\pm} = p_x \frac{1}{2m(x)} p_x - \mu \pm \beta \sigma_z + V \delta(x),
\end{eqnarray}
where $p_x$ is the momentum along $x$ direction and $V$ is the interfacial potential at the F/ISC interface $x=0$. $m(x)=m_F$ and $m(x)=m_S$ are the effective masses in the F and ISC regions, respectively. The wave functions $\psi(x)$ at the interface satisfy $\int_{0^-}^{0^+} H_{\pm}^{BdG} \psi(x) dx = \int_{0^-}^{0^+} E \psi(x) dx$. Because of the $\delta$ potential at the F/ISC interface and the mass difference between F and ISC, the derivative of $\psi(x)$ is discontinuous at $x=0$, which satisfies the conditions $\frac{\psi_S'(0^+)}{m_S}-\frac{\psi_F'(0^-)}{m_F}=\frac{2V}{\hbar^2}\psi(0)$ and $\psi_S(0)=\psi_F(0)\equiv \psi(0)$. Based on these boundary conditions, the LAR $R_A$ can be calculated.

\begin{figure}
\begin{center}
\includegraphics[width=8.0cm,height=5.0cm]{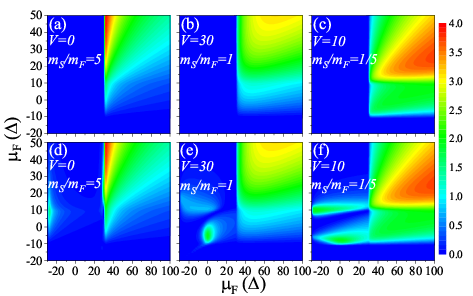} \\
\caption{ LAR coefficient $R_A$ versus potentials $\mu_F$ and $\mu_S$ at (a-c) $\theta=0$ and (d-f) $\theta=\pi/2$.}
\end{center}
\end{figure}

\begin{figure}
\begin{center}
\includegraphics[width=8.0cm,height=4.0cm]{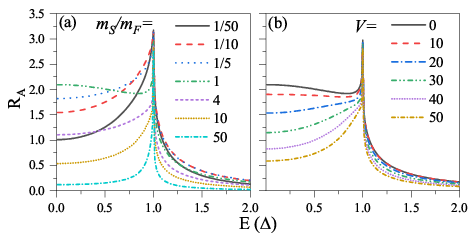} \\
\caption{ LAR coefficient $R_A$ versus incident energy $E$ with (a) $V=0$ and (b) $m_F=m_S$. Other parameters are set as $\mu_F=40$, $\mu_S=10$, and $\theta=\pi/2$.}
\end{center}
\end{figure}

The results on the LAR for F/ISC junction are shown in Figs. 13-14 for different values of the mass difference $m_S/m_F$ and the interfacial potential $V$, and other parameters are the same as these in Fig. 2, unless otherwise noted. One may clearly see that compared with that in Fig. 2(a), the normal LAR at $\mu_F>M$ is suppressed in some areas. However, the LAR in some special areas is enhanced [see Figs. 13(a) and 13(c)]. At $\theta=\pi/2$ in Figs. 13(d-f), the equal-spin AR could still occur in the spin polarized F with $\mu_F<M$, although it is suppressed by the large $m_S/m_F$ and $V$. Interestingly, for some suitable values of $m_S/m_F$ and $V$, such as $m_S/m_F=1/5$ and $V=10$ in Fig. 13(f), the areas for the equal-spin AR are broadened and changed. Fig. 14 presents the LAR as a function of incident energy $E$ for (a) the different mass differences and (b) the different interfacial potentials. For large values of mass difference [see Fig. 14(a)] or interfacial potential [see Fig. 14(b)], the LAR has a certain suppression. But the LAR coefficient $R_A$ can still keep the large values, e.g. larger than $1.01$ and $0.12$ at $E<\Delta$, even when the mass ratio $m_S/m_F$ reaches $1/50$ and $50$. Furthermore, the LAR is dramatically enhanced and has a sharp peak at $E=\Delta$, which is independent of $m_S/m_F$ and $V$, consistent with the previous result \cite{Blonder}. The above results demonstrate that the mass difference and the interfacial potential could quantitatively weaken the AR, but should have no essential effect on the occurrence of the AR.

\section*{Acknowledgments}

This work was supported by the NSFC (Grants No. 11974153 and No. 11921005), National Key R and D Program of China (Grant No. 2017YFA0303301), and the Strategic Priority Research Program of Chinese Academy of Sciences (Grant No. XDB28000000).

\end{CJK*}
\end{document}